\documentclass[11pt, a4paper,twocolumn]{article}
\usepackage{geometry}                

\usepackage{graphicx}
\usepackage{amssymb}
\usepackage{amsmath}
\usepackage{epstopdf}
\usepackage{fancyhdr}
\usepackage[authoryear]{natbib}
\usepackage{textcomp}
\usepackage{subfigure}
\usepackage{booktabs}
\title{Temperature profiles, plumes and spectra in the surface layer of convective boundary layers }
\author{Keith G. McNaughton\footnote{Kerikeri, New Zealand.
keith@mcnaughty.com} and  Subharthi Chowdhuri\footnote{Indian Institute of Tropical Meteorology, Pune, India}\\ \\ \\ \\For correspondence \quad keith@mcnaughty.com}

\begin{document}

\maketitle

\maketitle

\begin{abstract}
We survey temperature patterns and heat transport in convective boundary layers (CBLs) from the perspective that these are emergent properties of far-from-equilibrium, complex dynamical systems. We use the term `plumes' to denote the temperature patterns, in much the same way that the term `eddies' is used to describe patterns of motion in turbulent flows. We introduce a two-temperature (2T) toy model to define the cross-sectional areas of plumes, and so connect the scaling properties of temperature gradients, temperature variance and heat transport to this area. We then examine temperature ($T$) probability density functions and $w$-$T$ joint probability density functions, $T$ spectra and $wT$ cospectra observed both within and above the surface friction layer. Here $w$ is vertical velocity. We interpret these in terms of the frequencies at which each value is observed along a horizontal transect, and so with the cross-sectional areas of the plumes that give rise to them. In our discussion of $T$ spectra and $wT$ cospectra we focus first on the self-similarity property of the plumes and flux events above the SFL.  The CBL similarity parameters defined by McNaughton et al (Non-linear Processes in Geophysics 14, 257-271, 2007) are used throughout.  We interpret the $z^{1/2}$ dependence of the mixed length scale for wavenumbers in the $T$ spectra as reflecting the cross-sectional areas of the plumes, and so with the $z^{-1/2}$ form of the temperature profile, where $z$ is observation height. That is, mean temperature decreases with height as $z^{-1/2}$ because plume areas decrease as $z^{-1/2}$.  We introduce new scaling results for $T$ spectra and $wT$ cospectra from within the surface friction layer (SFL), based on a new analysis of data from the SLTEST experiment. We confirm earlier results showing that the scaling behaviours of $T$ spectra and $wT$ cospectra change for heights below $z/z_s<0.1$, where $z_s$ the height of the SFL, and come to display properties associated with random diffusion. We conclude by contrasting our interpretation of the role of buoyancy as a system-wide action in CBL flows with that of Richardson (Proc. Roy. Soc. London A 87, 354-373, 1920), who proposed a local action and whose ideas inform the current interpretation of the statistical fluid mechanics model of boundary-layer flows.

\end{abstract}

\section{Introduction}
\label{intro}
Turbulent transport in convective atmospheric boundary layers depends on the forms, sizes and energies of the patterns of motion---`coherent structures' and `eddies'---found within them. These patterns can be regarded as emergent properties of convective boundary layer (CBL) flows. CBL flows have much in common with flows in Rayleigh-B\'enard convection cells, particularly in windless conditions, but CBLs are open systems that grow during the daytime as warmer air is entrained through the capping inversion at their tops. Heat is therefore introduced through at both the upper and lower boundaries of CBLs, the eddies within them create patterns of temperature, or `plumes', which are also emergent properties of these flows. Our task is to describe these eddies and plumes, and their inter-relationships. The problem is a difficult one, and must necessarily be based on experiment since the governing equations cannot be solved. Though experimental observations and results from simulations have been analyzed in many ways, and much has been learned, the most systematic way to begin is by empirical scaling analysis allied with the concurrent development of a conceptual model. That is, we must learn to present these observations in universal ways. Success in finding appropriate scaling parameters will go hand-in-hand with understanding what controls the sizes of, energy and scalar concentrations in, and fluxes carried by particular classes of eddies and plumes. The forms of the eddies and plumes must be discovered by other means.

In previous works we have focussed on the scaling properties of spectra and cospectra, particularly of temperature ($T$) spectra and $wT$ cospectra from CBLs  \citep{McN07,Lau09,Cho19}, where $w$ is vertical velocity. The reason is that spectra naturally sort eddies by size, since the essential non-linearity of the governing Navier-Stokes equations denies linear superposition of eddies, so eddies of different kinds cannot coexist unless they are well-separated by size. Implicit in this argument is that only a small number of distinct  emergent patterns of motion can satisfy all the feedback requirements inherent in this flow system. Empirical results support this proposition \citep{Per75,Per86,McN07}.

The rationale for our basic set of scaling parameters is given in earlier papers \citep{McN04,McN07,Lau09}. The set has two fundamental length parameters: $\lambda$, representing the size of the largest turbulent structures in the CBL, and $z$, representing observation height. It has two parameters to describe the flow of mechanical energy through the CBL system: the outer dissipation rate, $\epsilon_o$, which is constant with height above the surface friction layer (SFL) \citep{Kai76,McN07}, and the dissipation velocity, $u_\epsilon$. The latter, with $z$, parameterizes the dissipation rate within the SFL. In practical terms, $\lambda$ is the peak wavelength of the streamwise velocity ($u$) spectrum, and $u_\epsilon=(kz\epsilon)^{1/3}$, where $\epsilon$ is the dissipation rate at height $z$ near the ground, and $k$ is the von K\'arm\'an constant. We also use the height of the SFL, $z_s$, given by 
\begin{equation}
z_s=\frac{u_\epsilon^3}{k\epsilon_o}
\label{z_sdef}
\end{equation}
The SFL is characterized by the presence of a special class of attached shear eddies. These develop upwards from the ground until their growth is terminated by interaction with detached eddies of similar size from the outer Kolmogorov cascade. Equation (\ref{z_sdef}) represents the height at which inner and outer dissipation rates match, as explained by \citet{McN04} and \citet{McN07}. Finally, we use the kinematic heat flux, $H=\left< w'\theta' \right>$ where the primes indicate deviations from the area means of $w$ and $\theta$, to construct a temperature scale. Buoyancy effects of the heat flux are accounted through the dissipation rates. 

What complicates the picture is that the basic length scales, $\lambda$, $z$ and $z_s$, appear not just alone but also in combinations, as mixed length scales and doubly-mixed length scales. A mixed scale is the geometric mean of two component length scales, so taking the form $\ell_1^{1/2}\ell_2^{1/2}$ where $\ell_1$ and $\ell_2$ are the component length scales. Doubly-mixed scales follow the same pattern, but with one of the component scales itself being a mixed scale. We observe that mixed scales are found when one kind of eddy or plume, with length scale $\ell_1$ exists within, and has sizes or aggregation properties organized by larger eddies with length scale $\ell_2$. It has been found that mixed and doubly-mixed scales often describe turbulence processes near smooth walls \citep[e.g.,][]{Alf84,DeG00,Met01,Bus09}, with half powers appearing in every case, even while dimensional consistency requires only that  a scale length takes the form $\ell_1^\alpha \ell_2^{1-\alpha}$, where $\alpha$ has any value. The possibility of mixed scales puts the search for suitable scales beyond the reach of simple dimensional analysis.

\begin{figure}[!tb]
\vspace{2mm}
	\begin{center}
		\includegraphics[width=.45\textwidth]{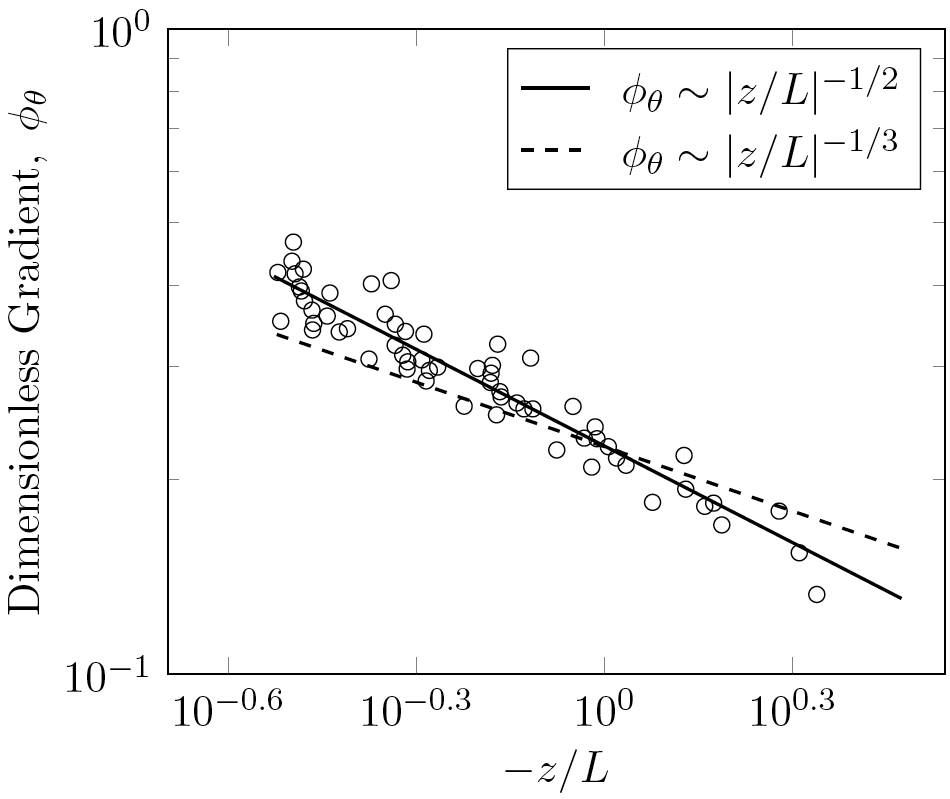}
		\caption{Dimensionless potential temperature gradient plotted against height, with each axis scaled according to the Monin-Obukhov similarity model so that $\phi_\theta=(zu_*/H) \partial \theta /\partial z$, where $u_*$ is the friction velocity and $z$ is scaled on the Obukhov length $-L$, as defined in (\ref{Olength}). The $-1/2$ power law describes the data over at least a decade of $-z/L$. Redrawn from \citet{Bus71}.}
		\label{Bprofile}
	\end{center}
\end{figure}

In this paper we survey the scales found to be successful in collapsing temperature ($T$) spectra and $w$-$T$  cospectra from above and within the SFL, and  interpret them in terms of the known properties of eddies and plumes found in CBLs. We include a discussion of the temperature profile in this survey, because its gradient follows a $-1/2$ power law for more than a decade, from mast top down to deep within the SFL, as shown in Fig.~\ref{Bprofile}. Power laws are of particular interest here because they indicate a scale invariance that is consistent with the self-similarity (collapse) of the $T$ spectra and $wT$ cospectra and so with a self-similarity of the underlying turbulent structures over that range. We find the half power law to be particularly interesting now that we have become familiar with mixed length scales for wavenumbers in our spectra and cospectra. Generally, we seek to explain the ensemble-averaged descriptions of these flows, as observed in experiments, in terms of the known properties of the underlying, emergent structures that exist in space and time in CBL flows. That is, we seek to extend our conceptual model of CBL turbulence and its transport properties beyond what we have attempted previously.

\section{Temperature gradients}
\label{sec:1}

We begin with mean temperature gradients in the atmospheric surface layer of CBLs since they are important in themselves, and they open an important window onto a self-similarity property of plumes near the ground. The temperature gradient in much of the surface layer of CBLs can be written as 
\begin{equation}
\frac{z u_*}{H}\frac{\partial \overline{\theta}}{\partial z}=- \alpha \left(\frac{z}{-L}\right)^{-1/2}
\label{KYprofile}
\end{equation}
using the scaling parameters proposed by the Monin-Obukhov similarity theory (MOST) \citep{Mon54}.  We use the term `surface layer' to mean the layer below the mixed layer in CBLs, or roughly the bottom 10\% of the CBL. Our usage is compatible with MOST usage, where the surface layer is defined as the layer where heat and momentum fluxes are sufficiently constant with height that they can be represented by their surface values throughout the surface layer. The SFL is a sub-layer of the surface layer in CBLs. In (\ref{KYprofile}) $u_*$ is the friction velocity, $H=\overline{w'\theta'}$ is the kinematic heat flux at the ground, $\overline{\theta}$ is time-averaged potential temperature, $\alpha$ is a constant and $L$ is the Obukhov length, defined by
\begin{equation}
-L=\frac{u_*^3 T_0}{kgH}
\label{Olength}
\end{equation}
where $g$ is the acceleration due to gravity and $T_0$ is surface temperature in degrees K. The $-1/2$ power in (\ref{KYprofile}) was established empirically and is not consistent with MOST itself, which predicts that the profile should approach a $-1/3$ power law in the limit of $u_*=0$ in windless convection, when $-z/L\rightarrow \infty$  \citep{Bus71}. The comparison is shown in Fig.~\ref{Bprofile}. We note that \citet{Kad90}, henceforth KY90, fit two segments of $-1/3$ power laws to their data from several sources, in conformity with their three-layer model of the surface layer and the requirements of MOST, even though the $-1/2$ power law provides a better fit over the full range $0.04<-z/L<4$ (KY90: Fig.1).  \citet{Fok83} report that the data from the Tsimlyansk  site in 1981 follow a $-1/2$ power law for all $-z/L>0.06$.

The scaling used in (\ref{KYprofile}) is not unique. Power laws indicate scale-free, self-similar behavior of the underlying structures giving rise to the power law, so temperature profiles can be scaled in whatever way best suits the context. Thus the scales $u_*$ and $-L$ in (\ref{KYprofile}) can be replaced by those that \citet{Lau09} found to give universal forms to $T$ spectra and $wT$ cospectra within the SFL. Using these scales we have    
\begin{equation}
\frac{z (z_s\epsilon_o)^{1/3}}{H}\;\frac{\partial \overline{\theta}}{\partial z}= -\alpha_1 \left(\frac{z}{z_s}\right)^{-1/2}
\label{Mprofile}
\end{equation}
where $\alpha_1$ is another positive constant. Here $u_\epsilon=(kz\epsilon)^{1/3}$ and is the value used to calculate $z_s$ by (\ref{z_sdef}). The dissipation rate $\epsilon$ is measured deep within the SFL, so $u_\epsilon$ is like $u_*$ but includes the effect of variations in surface shear stress on dissipation rates near the ground, as first discussed by \citet{Tow61}. Unlike $L$, $z_s$ does not go to zero in windless convection. 

We now have two equations, both describing the same temperature profile but scaled in two different ways.  Both can achieve some success if the parameters of the two models are statistically correlated. Results from SLTEST \citep{Cho19} do show a strong correlation between $z_s$ and $L$. If this is generally true then both (\ref{KYprofile}) and (\ref{Mprofile}) can be expected to give fairly good accounts of observed temperature profiles, despite their very different foundations. The question is which scaling scheme gives less scatter and the more `universal' results? 

We await direct experimental comparisons, but in the meanwhile we find it more convenient to work with (\ref{Mprofile}) because its parameters are related to the geometrical properties of particular classes of eddies and plumes in CBLs, and it is these properties that we investigate here. Before starting we rewrite  (\ref{Mprofile}) as
\begin{equation}
\frac{\partial \overline{\theta}}{\partial \zeta}= -\alpha_1 \theta_* \zeta^{-3/2}
\label{M2profile}
\end{equation}
where $\zeta=z/z_s$ is a dimensionless length, $\theta_*$ the temperature scale $H/(z_s\epsilon_o)^{1/3}$ and $\alpha_1$ a constant.  

An interesting question is the plausibility of the profile forms (\ref{KYprofile}) and (\ref{Mprofile}) above the SFL.  KY90 discuss a three-layer model of the lower CBL, and they find it surprising that (\ref{KYprofile}) applies in their convective sublayer where $u_*$ is not a relevant parameter, reasoning that  since $L$ depends on $u_*$ then $L$ should have no significance there. Equally, our profile expressions, (\ref{Mprofile}) and (\ref{M2profile}), employ the length scale $z_s$, which we would not expect to be relevant above the SFL. In our case the problem is resolved when we note that the height variable in (\ref{M2profile}) can be rescaled by replacing $z_s$ with the outer length scale, $\lambda$. We can then redefine the scaled height variable as $\zeta=z/\lambda$, and redefine the temperature scale as $\theta_*=H/(\lambda \epsilon_o)^{1/3}$ in (\ref{M2profile}). With these we can rewrite (\ref{M2profile}) in a form identical to (\ref{M2profile}) but with $\alpha_1$ redefined as $\alpha_2=\alpha_1(z_s/\lambda)^{1/6}$. In this new version of  (\ref{M2profile}) the length scale $\lambda$ reflects the streamwise extent of the largest eddies found in CBLs, so it is closely related to the height of the temperature inversion layer that caps, and so marks the vertical height of CBLs, $z_i$. The ratio of $z_i$ to $\lambda$ will depend on the aspect ratio of the large cellular or roll structures in the CBL, and this relationship has not been fully explored. We note that \citet{Kai76} reported that $\lambda=1.5z_i$ for thirteen runs from the five days before a lightening strike terminated the Minnesota experiment. Overall, we see that (\ref{M2profile}) can be written in either inner-scaled or outer-scaled versions by choosing to set either $\eta=z/z_s$ or $\eta=z/\lambda$, respectively.

KY90 also present profiles of the standard deviation of temperature, $\sigma_\theta$, observed during many experiments, both within and above their `convective-dynamic sublayer'. The top of this sublayer corresponds well to the top of our SFL, but our SFL extends right down to the ground. We will return to this later, but here we note that KY90 plot their profiles within and above the SFL in two segments using MOST scaling parameters, and find that
\begin{equation}
\frac{u_*\sigma_\theta}{H} = \beta \left(\frac{z}{-L}\right)^{-1/3}
\label{KYsdT}
\end{equation}
in each, though the constant $\beta$ has a different value in each segment. The $-1/3$ power is consistent with expectation from MOST in the limit of free convection and with the earlier results of \citet{Pri60,Bus71} and \citet{Wyn72}. This behavior extends down to $-z/L\approx 0.05$. The side-step in the profile of KY90 is unexplained.

Once again, this profile can be written in terms of our favored parameters, giving 
\begin{equation}
 \sigma_{\theta}= \beta_1\theta_* \zeta^{-1/3}
\label{MsdTA}
\end{equation}
where $\zeta=z/\lambda$.
This profile agrees with the variances found by integrating the areas under scaled temperature spectra observed above the SFL \citep{McN07,Cho19}, and also within it according to the new $T$ spectra presented below, which self-similarity extends down to similarly low levels. 

To summarize: temperature profiles follow a $z^{-1/2}$ power law through most of the surface layer, while profiles of its standard deviation, $\sigma_T$, follows a $-1/3$ power law over a similar range. Since both profiles follow power-laws we infer that these profiles are associated with some kind of scale-free, self-similar structures. In the lowest part of the SFL the profiles depart from these power law forms and the temperature profile follows a log law while the $\sigma_\theta$ profile becomes constant with height. These behaviors indicates that heat transport there is effected by another kind of self-similar eddy structure. 

In searching for the mechanisms that underly these mean profiles we will appeal to the geometric and scaling properties of various kinds of plumes, and to do this we must have a working understanding of the nature of plumes.

\section{Modeling plumes}

We use the word `plume' to mean any connected volume of air defined by its temperature, so our usage goes beyond the narrow regions of ascending air that common usage implies. Plumes may frequently take this form, but we include all warm and cool updrafts and downdrafts, independent of their shape or size. We restrict our discussion to heights above any viscous sublayer or roughness sublayer immediately above the ground, and to below the mixed layer where mean temperature becomes almost constant with height. That is, we deal with the  atmospheric surface layer as it is usually defined in boundary-layer meteorology.

Plumes, like eddies, defy exact definition. They are recognizable `patterns of scalar concentration', just as eddies are `patterns of motion', and they can have discoverable scaling properties even while the patterns themselves are more-or-less ill-defined. There is an intuitive connection between a plume or an eddy and the concept of a strange attractor, since the actual shapes of plumes and eddies will exhibit many variations on a theme. Since we find the concept of plumes, and the shapes of plumes to be useful concepts, we now introduce a toy model in which plume boundaries, and so plume shapes and cross-sectional areas, can be rigorously defined.

\subsection{A two-temperature plume model}

Our 2T model is a toy model that allows temperatures to have just one of two values at each point while allowing the velocity field to have all of its observed properties. This is a great simplification of real temperature fields, and it does create some problems, but this 2T model also leads to some simple relationships that will help us to understand the similarity properties of real plumes observed in real boundary layers.

We consider a plane horizontal flow in which there are two kinds of air, each distinguished by its potential temperature, $\theta_u$ or $\theta_d$. Each kind is organized into localized regions called plumes.  Warmer plumes originate somewhere below and move upwards on average, while cooler plumes originate somewhere above and generally move downwards. We neglect molecular diffusion, so the up-and down-plumes retain the distinct identities imparted to them at their sources, even while they may be stretched and folded in complicated ways. Warm plumes move upwards with a mean velocity $\left <w_u\right>$, and cool plumes move downwards at $\left<w_d\right>$, where the angle brackets indicate means over the areas of each kind of plume measured on a horizontal plane at a chosen height $z$.  A heat flux is maintained by air being converted from $\theta_d$ to $\theta_u$ in a thin layer adjacent to the ground, at heights below those of interest here. In the real world this heating would occur within a viscous sub-layer or roughness sub-layer adjacent to the ground, which layers are excluded from  present consideration.

We will use the term `up-plume' to indicate a parcel of air whose temperature is set near the ground and which moves generally upwards over time, and the term `down-plume' for a parcel of air whose temperature is set above and moves generally downwards. We note that the up- and down- designations refer only to their average motions, not the local motion at any particular place and time, which may be in either direction.

\subsection{Relationships in the 2T model}

For simplicity we assume that air density is constant, independent of temperature. The 2T model then leads to a number of basic relationships.

\subsubsection{Continuity}

Let the warmer and cooler plumes occupy fractional areas $f_u$ and $f_d$ on any horizontal plane at height $z$. These fractions comprise the whole, so
\begin{equation}
f_u+f_d=1
\label{areas}
\end{equation}
The condition that there be no net vertical transport of air is written as 
\begin{equation}
\left<w_u\right> f_u+\left<w_d\right> f_d=0
\label{dryair}
\end{equation}
where, for simplicity, we have neglected the effect of temperature on air density.

\subsubsection{Mean temperature gradient}

The numbers of $\theta_u$ and $\theta_d$ values found in a sample taken along a horizontal transect will be proportional to the area fractions of up- and down-plumes, so the mean temperature on the horizontal plane at any value of $z$ is
\begin{equation}
\left<{\theta}\right>=f_u \theta_u+f_d \theta_d
\label{tmean}
\end{equation}
Using this with  (\ref{areas}) we can write the mean temperature gradient as
\begin{equation}
\frac{\textrm{d}\left<{\theta}\right>}{\textrm{d} z}=(\theta_u-\theta_d) \frac{\textrm{d} f_u}{\textrm{d} z}
\label{tgrad1}
\end{equation}
or its twin 
\begin{equation}
\frac{\textrm{d}\left<{\theta}\right>}{\textrm{d} z}=-(\theta_u-\theta_d) \frac{\textrm{d} f_d}{\textrm{d} z}
\label{tgrad1b}
\end{equation}
We notice that  the temperature difference $(\theta_u-\theta_d)$ is a constant, so the temperature gradient depends solely on how the fractional area of up-plumes d$f_u/$d$z$, or its complement d$f_d/$d$z$, changes with height.  If warm plumes become faster and thinner as they rise then area-mean temperature will decrease with height.

\subsubsection{Plume mean velocity variance}

Because there are two mean vertical velocities  there will be a plume mean velocity variance. As for temperature, the numbers of $w_u$ and $w_d$ values selected in a given sample will be in proportion to their area fractions, $f_u$ and $f_d$, at a given height.  Starting with 
\begin{equation}
\sigma_{<w>}^2=\left<{\left<w\right>^2}\right>-\left<{\left< w\right>}\right>^2\\
\end{equation}
we have
\begin{align}
\sigma_{\left<w\right>}^2=&f_u \left<w_u\right>^2+f_d \left<w_d\right>^2\nonumber\\
& -(f_u \left<w_u\right>+f_d \left<w_d\right>)^2
\end{align}
giving
\begin{equation}
\sigma_{\left<w\right>}^2=f_u f_d \; (\left<w_u\right>-\left<w_d\right>)^2
\label{vvar1}
\end{equation}
which can also be written as
\begin{equation}
\sigma_{\left<w\right>}^2=-\left<w_u\right>\!\left<w_d\right>
\label{vvar2}
\end{equation}
by using (\ref{dryair}).

We note that (\ref{vvar2}) gives the variance of the plume mean velocities, not the total variance of vertical velocity, $\sigma_w^2$ . This is an important distinction, and we have no guarantee that $\sigma_w$ and $\sigma_{\left<w\right>}$ will be equal, or even that they will scale the same way.

\subsubsection{Temperature variance}

Because the up- and down- plumes have different temperatures there is temperature variance on the $(x,y)$ plane. Starting with 
\begin{equation}
\sigma_\theta^2=\left<{\theta^2}\right>-\left<{\theta}\right>^2\\
\end{equation}
we write
\begin{equation}
\sigma_\theta^2=f_u\theta_u^2+f_d\theta_d^2-(f_u\theta_u+f_d\theta_d)^2
\end{equation}
and so
\begin{equation}
\sigma_\theta^2=f_u f_d \; (\theta_u-\theta_d)^2
\label{tvar1}
\end{equation}
Thus $\sigma_\theta^2$ is proportional to $f_uf_d$.  We note that $(\theta_u-\theta_d)$ is independent of height, so it may be regarded as the temperature  scale in our 2T model. The product of the fractional areas, $f_u f_d$, will depend on height scaled by an appropriate height scale.

\subsubsection{The heat flux}

The  kinematic heat flux is given by
\begin{equation}
\left<w'\theta'\right>=\left<w_u\right> f_u \theta_u+\left<w_d\right> f_d \theta_d
\label{heatflux}
\end{equation}
where $\left<w'\theta '\right>$ is averaged over the whole horizontal plane. We will use the symbol $H$ for this heat flux. Combining (\ref{heatflux}) with (\ref{dryair}) gives
\begin{equation}
H=\left<w_u\right> f_u( \theta_u- \theta_d)
\label{heatflux1}
\end{equation}
and its twin 
\begin{equation}
H=-\left<w_d\right> f_d( \theta_u- \theta_d)
\label{heatflux2}
\end{equation}
Since $H$ and $( \theta_u- \theta_d)$ are both constant with height we have an inverse relationship between $\left<w_u\right>$ and $f_u$ and between $\left<w_d\right>$ and $f_d$. On average the up-plumes get thinner and faster as they ascend, while the down-plumes get thinner and faster as they descend. 

Multiplying left and right sides of the last two equations together we get
\begin{equation}
H^2=-\left<w_u\right>\!\left<w_d\right>f_u f_d( \theta_u- \theta_d)^2
\label{heatflux3}
\end{equation}
which can be rewritten as
\begin{equation}
H^2=\sigma^2_{\left<w\right>} \sigma^2_\theta
\label{heatflux4}
\end{equation}
using (\ref{vvar2}) and (\ref{tvar1}). The temperature variance scales on $H^2/\sigma^2_{\left<w\right>}$.

\subsection{Limitations of the 2T model}

The 2T model is a toy model, designed to help us to understand the link between plume geometry and mean vertical profiles of temperature and temperature variance. Its principal value is that it allows us to talk of the areas of plumes without worrying too much about how the boundaries of those plumes should be defined. However, the 2T approximation has its limitations. 

We can explore these by considering the temperature signal along a horizontal transect, as represented by the 2T model, and comparing this with a more-realistic representation.  A plot of temperature along a 2T transect would look like a comb with many teeth, with a bases at $\theta_d$ and all of the same height, $\theta_u$, but with various widths and grouped into clusters of various sizes and compositions. The total lengths of the teeth and the gaps would be in ratio $f_u/f_d$ along a representative transect. A more realistic model would allow molecular diffusion to operate, rounding the corners of the teeth, thereby reducing the heights of smaller teeth. Despite these changes the resulting groupings of fuzzy plumes would have almost the same mean temperature as their 2T originals. On the other hand the temperature variances would be reduced. For this reason we can have some confidence in (\ref{tgrad1}) but we must be wary of (\ref{tvar1}) and relationships that follow from it. 

Despite its limitations, the 2T model allows us to talk of the lengths and areas of plumes, and so to interpret the power-law temperature profile geometrically, in terms of the self-similar tapering with height of plume cross sections. We believe that this remains useful even while real plumes are blurry composites of whole populations of smaller plumes, and even while the plumes originating from near the ground and the top of the CBL have, in reality, a wide range of temperatures. Also, the 2T model allows the profiles of $\partial \overline{\theta} / \partial z$ and $\partial \sigma_\theta /\partial z$ to depend differently on height, since $\partial \overline{\theta} / \partial z \propto d f_u/d z$, from (\ref{tgrad1}), while the second $\partial \sigma_\theta /\partial z \propto d( f_u^{1/2}f_d^{1/2})/d z$, from (\ref{tvar1}). Observations show the profiles exhibit different powers of $z$, even while this is contrary to the dictates of MOST above the SFL, in what Tennekes called the "local free convection layer" \citep{Ten70}. We will adopt  the 2T model as a useful, qualitative guide as we seek to understand the mechanisms that underly the observed, ensemble-mean properties of temperature and heat transport in the surface layer.

Before using these ideas to interpret real observations we extend our terminology to include composite plumes. We have used the terms `up-plume' and `down-plume' to designate plumes that originate either near the ground or far overhead, so the up- and down prefixes describe their mean direction of movement even while their local motion might be either direction at any particular time and place. As we move towards a more realistic representation of reality we will add the idea of composite plumes. These are defined as aggregations of up- and down-plumes having an identifiable common boundary and local bulk velocity. They are compatible with the 2T model, in that the up- and down- identities of their parts remain intact, but they will have length scales and concentration scales that reflect the common properties imparted to them by the rest of the flow, which properties lie beyond the scope of the simple 2T model.

\section{Experimental}

In this paper we focus on observations made during the SLTEST experiment, conducted in the Great Western Desert of Utah in 2005. The SLTEST site was close to an ideal site for our experiment. Nine sonic anemometers were mounted on a mast, facing North and spaced logarithmically over an 18-fold range of heights, from 1.42 m to 25.7 m. The fetch area was a flat, unobstructed playa surface stretching for 100 km to the North. The surface itself was particularly smooth during the experiment in May and June of 2005, since frequent rain in the winter and spring had eliminated the usual surface cracking and suppressed the growth of salt crystals, giving a very uniform surface of finely dispersed clay. This reduced the surface drag and resulted in SFLs that were often only a few meters deep, so, with winds predominantly from the North, the experiment yielded an unexpectedly large number of useable observations from above the SFL. On other occasions the SFL was deeper so the instruments recorded profiles within the SFL, as was envisaged in the original experimental design.  Further information is given by \citet{McN07}.

\section{Plumes above the SFL}

The $T$ and $w$ signals contain a great deal more information than just their means, standard deviations and total heat fluxes, and this information can be used to explore the properties of plumes and heat-flux events. A probability density plot of temperature, for example, gives information on the relative areas occupied by plumes of various temperatures, while $T$ spectra and $wT$ cospectra give information on the lengths of plumes and flux events, and on their scaling properties. We now look at such information, starting with observations from above the SFL, where $u_\epsilon$ is not a relevant parameter because it represents the velocity scale of the shear eddies within the SFL. Our set of similarity parameter then reduces to  \{$z, \lambda, \epsilon_o, H$\}.

\subsection{$T$ and $wT$ probability distributions}

\begin{figure}[!tbp]
\vspace{2mm}
	\begin{center}
		\includegraphics[width=.47\textwidth]{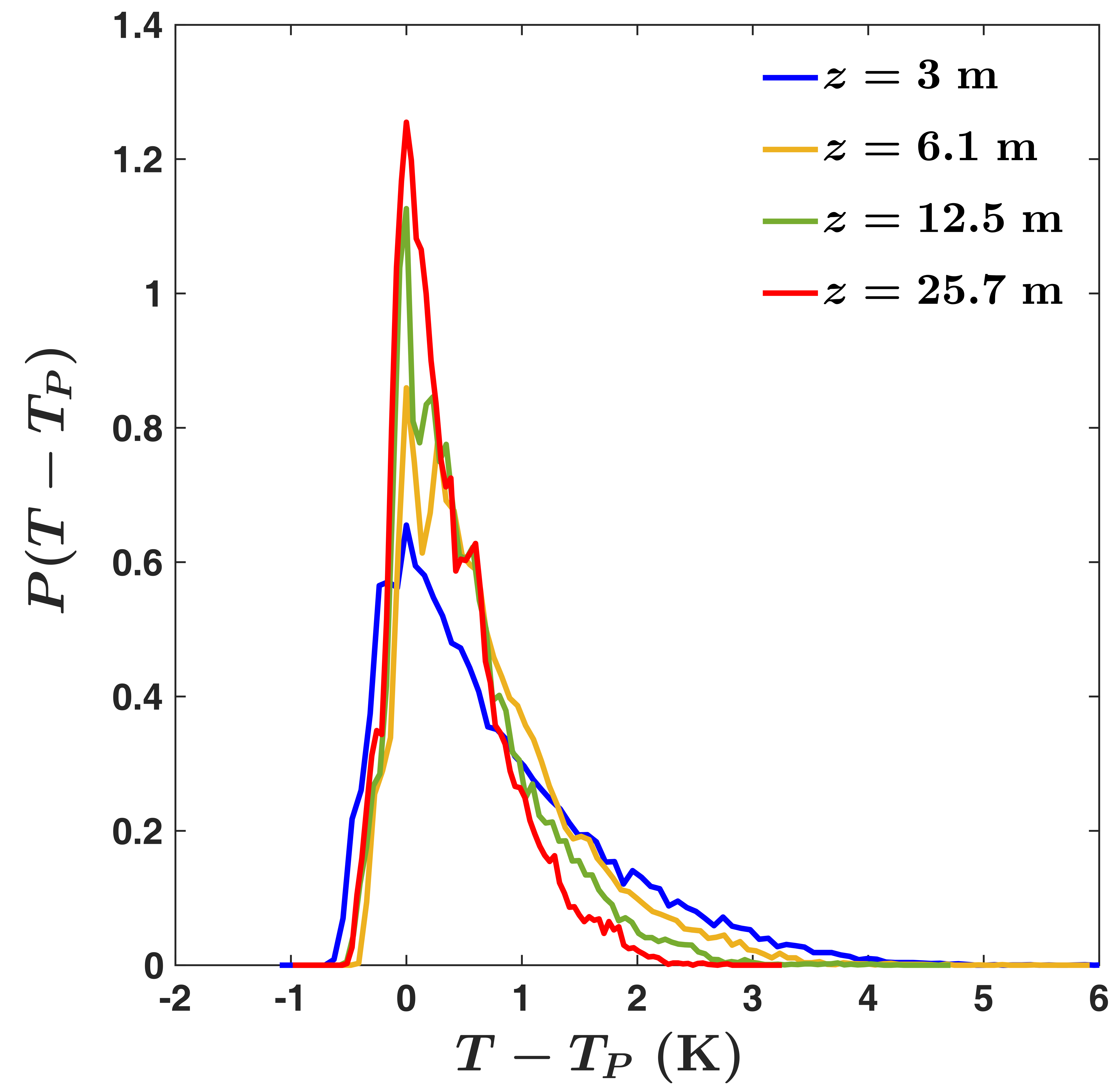}
		\caption{Probability distributions of temperatures at four heights above the SFL site at 2100-2130 on 24 May 2005, when $\lambda=1230$ m. Here $T_P$ is the temperature with peak probability. The distribution is strongly skewed towards cooler temperatures because cooler, subsiding air occupies the greater fraction of total area. Skewness increases with height as the area fractions change: the area of cooler subsiding air increases while that of warmer rising air increases. }
		\label{T_pdf_largez}
	\end{center}
\end{figure}

\begin{figure}[tb]
\vspace{2mm}
	\begin{center}
		\includegraphics[width=.47\textwidth]{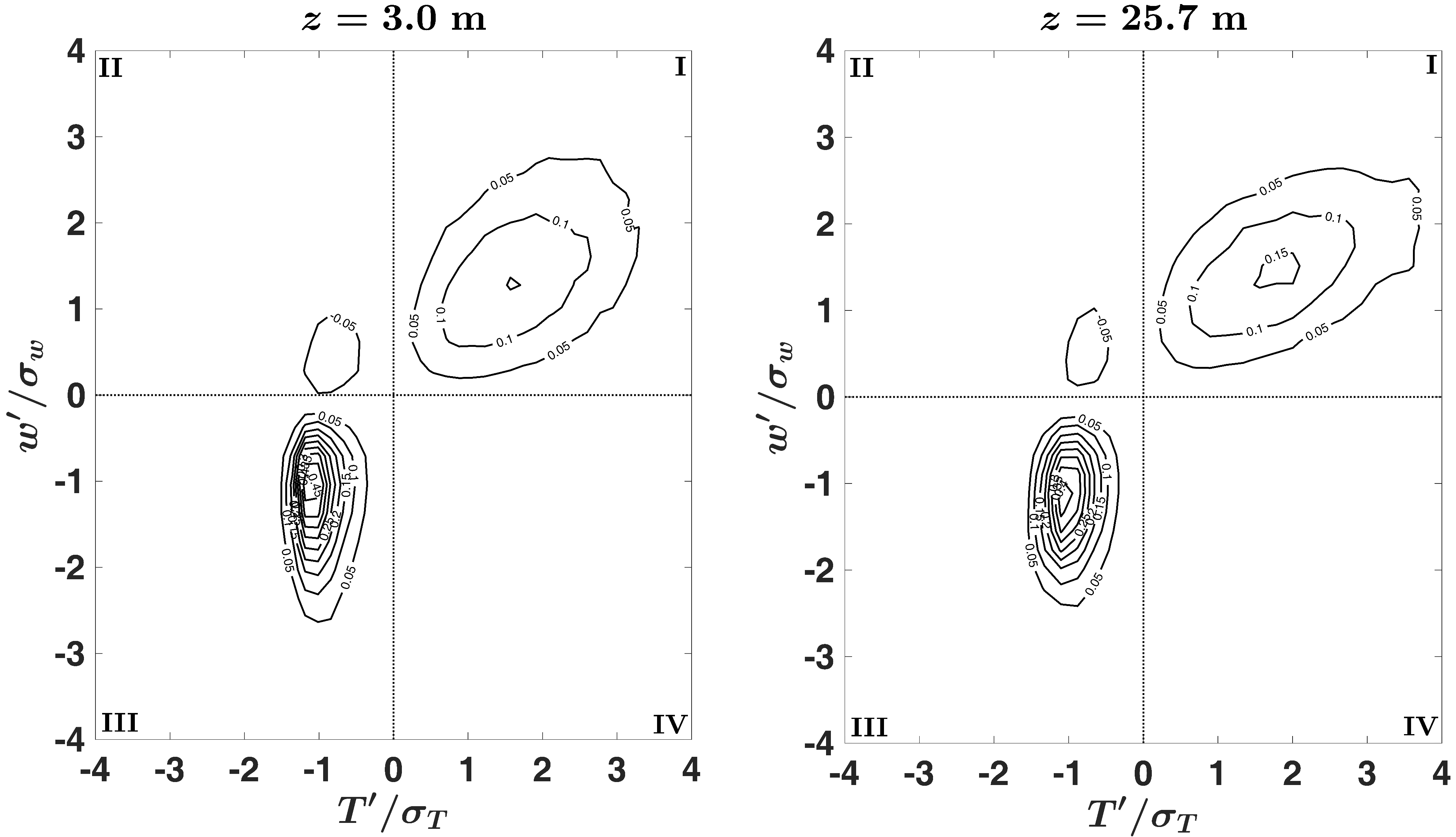}
		\caption{Joint $w'$-$T'$ distribution at 3.0 m and 25.7 m plotted on the $w'/\sigma_w, T'/\sigma_T$ plane and averaged over a selection of runs when both levels were above the SFL and 900 m $<\lambda < $ 1300 m. The primes indicate deviations from the mean at each level. Heat transport is highly organized, being dominated by warm updrafts (I) and cool downdrafts (III) with little reverse transport in quadrants II and IV.}
		\label{wT_flux_largez}
	\end{center}
\end{figure}

Fig. \ref{T_pdf_largez} shows temperature PDFs at four heights, all above the SFL during a particularly unstable run at SLTEST. These are based on time-series data measured by fixed instrument. To interpret them we adopt Taylor's frozen turbulence hypothesis, by which the results are equivalent to observations taken at equal steps along a streamwise transect, and so represent the relative areas occupied by air of each temperature. Fig.~\ref{T_pdf_largez} shows that temperatures are not limited to just two values, as the 2T model would have it, but have a continuous range of temperatures.  Even so, we can retain the general idea of up- and down-plumes by allowing real up- and down-plumes to have a range of source temperatures. This being so, down-plumes will mostly be represented on the left of the diagram and up-plumes on the right. We notice that probabilities increase with height for $T-T_P<1$ and decrease with height for $T-T_P>1$. That is to say, these empirical results support the interpretation that up-plumes occupy an decreasing area fraction as height increases while the area fraction of down-plumes increases. This is consistent with what is suggested by (10), from the 2T model, and with the general observation that temperature decreases with height in the surface layer of CBLs. 
 
The joint $w-T$ pdfs shown in Fig. \ref{wT_flux_largez} supports this interpretation. They show the distribution of  the heat flux on the $w'$ and $T'$ axes at 3.0 m and $25.7$ m averaged over a selection of runs when both levels were above the SFL. It shows the joint probability distribution of $w'$ and $T'$ weighted by the product $w'T'$ at each point. The contributions of warm ejections (quadrant I) and sweeps plumes (quadrant III) are quite distinct, and very little heat is transported in the `wrong' direction, in quadrants II and IV. Heat transport is therefore highly organized above the SFL. 

In making these arguments we must accept that the frozen turbulence hypothesis is not necessarily correct because eddies are not simply carried along in the flow, as \citet{Tay38} would have it, but the phase velocities of larger eddies and plumes can be faster than the local mean velocity in wall-bounded shear flows. Even so, the features of Fig.~\ref{T_pdf_largez} that we rely on for our interpretation would remain valid were a more exact representation of area-sampled pdfs available.

\subsection{$T$ spectra and $wT$ cospectra above the SFL}

The mean profiles and PDFs considered so far tell us nothing about the lengths of plumes. For that we turn to the analysis of $T$ spectra and $wT$ cospectra. The surface layer above the SFL is a good place to begin because the power-law forms of the $T(z)$ and $\sigma_\theta$ profiles above the SFL imply scale-invariant properties for the underlying plumes there. \citet{McN07,Cho19} have identified scales that collapse the $T$ spectra and $wT$ cospectra observed over a range of conditions onto `universal' curves, so demonstrating self-similarity in the properties of the plumes that give rise to them. These scales provide information on what controls the behavior of the underlying plumes. In this they followed along the path pioneered by \citet{Per75}, who studied eddies using these techniques.

\citet{McN07,Cho19} found that $T$ spectra and $wT$ cospectra observed above the SFL collapse onto universal curves in three wavenumber ranges, each with its own scales. We will call these the small-, mid- and large-wavenumber ranges.  The small-wavenumber range extends up to $\kappa \lambda \sim 2\pi$ and the small-wavenumber range down to $\kappa z \sim 2\pi$. Spectra in the the mid- and large-wavenumber ranges appear to be universal across sites, while the small-wavenumber spectra are somewhat different at the very uniform SLTEST site and the more heterogeneous site of the CAIPEEX experiment in India \citep{Cho19}. This lack of universality among sites is consistent with the well-known differences observed for the largest flow structures in different flows, such as pipe, channel and boundary layer flows \citep{Mon09}. It seems that the largest flow structures are influenced by the boundary conditions, particularly as they influence the overall flow geometry, while the mid- and small-scale structures depend on internal dynamics only and so have universal properties.  

Here we review the scaling results of  \citet{McN07,Cho19} and extend our interpretation of them. The scales found to collapse the $T$ spectra and $wT$ cospectra in each of the three ranges at SLTEST are given in Table \ref{ASFL}. The spectra themselves are given in the original reports. A novel feature of the length scales is that many of them are mixed scales, or even doubly-mixed scales. Here we adopt a qualitative explanation for mixed scales based on our general understanding that eddies of all scales in the CBL are parts of a single, coordinated flow system. Our hypothesis is that a mixed length scale $\ell_1^{1/2}\ell_2^{1/2}$ implies that a larger-scale eddy process, with length scale $\ell_1$, `organizes'  smaller plumes with length scale $\ell_2$. For the word `organizes' we have in mind such processes as sweeping collections of up-plumes together into composite plumes. Doubly-mixed length scales are interpreted similarly, with $\ell_1^{1/2}\ell_2^{1/2}$-scale eddies organizing smaller plumes with $\ell_3$-scale. We will apply this interpretation to the empirical scaling results.

\subsubsection{Mid wavenumbers} 
\label{midplumesA}

We start with mid-wavenumber ranges of $T$ spectra and $wT$ cospectra because they account for most of the $T$ variance and $wT$ fluxes observed throughout the surface layer, and they are found to be universal across sites. The mid range includes the peak regions of these spectra and cospectra. More practically, the search for the scales able to collapse mid-range spectra is made simpler because peak wavenumbers can be scaled unambiguously before going on to choose variance and covariance scales. These procedures are described by \citet{McN07,Lau09}, and the identified scales are shown in Table \ref{ASFL}. 

A notable feature of the scales recorded in Table \ref{ASFL} is that many of them are mixed scales, so that the properties of spectra depend on the ``outer'' length scale $\lambda$ even quite near the ground. This is in good qualitative agreement with the DNS simulation results of \citet{Fod19}, who found Monin-Obukhov similarity theory inadequate when describing plume properties and concluded that ``updraft properties are not just determined locally, but also by outer scales''. Our empirical studies of the scaling properties of spectra confirm that observation. We now turn to a more detailed interpretation of these scales.

\begin{table*}
\caption{Length and variance scales of $T$ spectra and $wT$ cospectra measured above the surface friction layer. This table summarizes the scales identified by \citet{McN07} and \citet{Cho19}. }
\label{ASFL}
\centering
\begin{tabular}{@{}llll@{}} 
\hline \\
Scales \rule{15mm}{0mm}		& \parbox{2.4cm}{Small \qquad wavenumbers}	& \parbox{2.4cm}{Mid \qquad wavenumbers}	& \parbox{2.4cm}{Large \qquad wavenumbers}		\\

&&&\\
\hline
$\kappa S_{ww}$&&&\\
\quad length		&$\lambda$	&$z$	&$z$\\
\quad variance  \rule{0pt}{18pt}	 &\( \displaystyle  (\lambda\epsilon_o)^{2/3}\left( \frac{z}{\lambda}\right)^{4/3}  \)	&$(z\epsilon_o)^{2/3}$	&$(z\epsilon_o)^{2/3}$\\

&&&\\

$\kappa S_{uu}$, $\kappa S_{vv}$	&&&\\
\quad length  	&$\lambda$	&$\lambda$	&$\lambda$\\
\quad variance \rule{0pt}{18pt}		& $(\lambda\epsilon_o)^{2/3}$	&$(\lambda\epsilon_o)^{2/3}$	&$(\lambda\epsilon_o)^{2/3}$\\

&&&\\
$\kappa S_{TT}$	&&	&\\
\quad length 	&$\lambda$	&$\lambda^{1/2}z^{1/2}$	&$\lambda^{1/4}z^{3/4}$\\
\quad variance  \rule{0pt}{18pt} 	&\( \displaystyle H^2(\lambda\epsilon_o)^{-2/3}\left( \frac{z}{\lambda}\right)^{-1/3}  \)	&\( \displaystyle H^2(z\epsilon_o)^{-2/3}  \)		&$H^2(z\epsilon_o)^{-2/3}$\\

&&&\\

$\kappa S_{wT}$	&&&\\
\quad length	&$\lambda$	&$\lambda^{1/4}z^{3/4}$	 &$z$\\
\quad covariance	 \rule{0pt}{18pt}	&\(  \displaystyle H\left( \frac{z}{\lambda}\right)^{1/2}   \)	&\( \displaystyle H \left( \frac{z}{\lambda} \right)^{1/12} \)  	&$H$\\

&&&\\ \hline

\end{tabular}

\end{table*}

The $T$ variance scale is $H^2(z\epsilon_o)^{-2/3}$, which implies a plume velocity scale $(z\epsilon_o)^{1/3}$. This is also the $w$ variance scale, and it identifies impinging outer Richardson eddies as the eddies active in shaping the plumes at this scale \citep{McN07}. The length scale here is the mixed length scale $\lambda^{1/2} z^{1/2}$, which we interpret as signifying an interaction between the plumes created by the impinging outer Richardson eddies and the larger, $\lambda$-scale eddies whose lower parts extend down into the surface layer. We need more information before we can go further, so we delay further interpretation until the end of this section. However, we do note that the peaks of the  $T$ spectra lie at $\kappa\lambda^{1/2}z^{1/2}=5.5$ \citep{Cho19}, so peak plumes are long in the streamwise direction, and they elongate further as height increases. This is consistent with the shapes of sheet plumes that comprise the rising walls of warm air that converge towards the bases of the thermals at the top of the surface layer.

We note that the $T$ variance scale, $H^2(z\epsilon_o)^{-2/3}$, is not the scale from the 2T model,  $(\theta_u-\theta_d)$, since the latter is independent of $z$. The difference lies in the identity of the plumes above the SFL. The properties of 2T up- and down-plumes are set at source, and their identities are preserved as they rise or sink, on average, thereafter, however much stretched, folded and tangled they may become. The lengths identified by the position of the spectral peaks, on the other hand, represent the aggregation properties of plumes as well as their individual properties.  As a result, the plumes whose temperature scales on $H^2(z\epsilon_o)^{-2/3}$ are composite plumes with characteristic temperatures that decrease with height as more and more clear air (down-plume) is added to the composite. Only if there were no dissipation and we could focus on just the original filaments of up-plumes would the variance scale be $(\theta_u-\theta_d)$.

The $T$ variance scale can also be written as  $H(z_s\epsilon_o)^{-1/3}$$\left(z/z_s\right)^{-1/3}$, and  as $H(\lambda\epsilon_o)^{-1/3}\left(z/\lambda\right)^{-1/3}$. The first of these is the temperature scale appropriate for plumes within the SFL, but modified by the height function $(z/z_s)^{-1/3}$ to express the diminishing the fraction of `original' up-plume air as height increases. The second introduces the temperature scale of the thermals at the top of the surface layer, $H(\lambda \epsilon_o)^{-1/3}$ times the height function $(z/\lambda)^{-1/3}$, which represent the same aggregation process towards the final temperature of the thermals. This flexibility of interpretation reflects the scale-free nature of the process itself: the whole up-plumes are embedded in regions of the flow that converge laterally and accelerate upwards, and the shapes of the embedded original plumes within the composites will be similar to the envelope shapes of the composite plumes. This self-similar property is consistent with the length-scale-independent, power-law form of the mean temperature profile.

We can go further by focussing on the first interpretation: that the plume properties reflect the action of impinging outer Richardson eddies, which sweep together larger and larger aggregates of the original plumes rising from within the SFL. If these impinging outer Richardson eddies acted in isolation from eddies of other kinds then plume lengths would scale on $z$ alone, rather than on $\lambda^{1/2}z^{1/2}$. The difference can be attributed to the up-plumes all being embedded in regions of the flow where the horizontal flow convergence, $\partial u/\partial x+\partial v/\partial y$, depends on $(z/\lambda)^{-1/2}$. The absence of heat flux in quadrant IV of Fig.~\ref{wT_flux_largez} tells us that the up-plumes are all associated with the updrafts created by these convergences, so we can directly associate the $(z/\lambda)^{-1/2}$ power law of the convergences with a similar dependence of the total area of the up-plumes. The length scale displayed by the mid-range of the $T$ spectrum is consistent with the $-1/2$-power law followed by the mean temperature profile.

Scales for $wT$ cospectra are more complicated. The lengths of $wT$ flux events scale on the doubly-mixed scale $\lambda^{-1/4}z^{-3/4}$ (because wavenumbers scale on $\lambda^{1/4}z^{3/4}$), Table~\ref{ASFL}). Peaks of $wT$ cospectra are found to be at $\kappa \lambda^{1/4}z^{3/4}=1.2$, while the wavenumbers of the peaks of $T$ spectra scale differently and are at $\kappa \lambda^{1/2}z^{1/2}=5.5$  \citep{Cho19}. Peak wavenumbers for $wT$ flux events are therefore from 1.5 to 2.5 times larger than those for $T$ plumes as $z/\lambda$ increases from 0.01 to 0.1. Flux events are therefore shorter than plumes, so we must conclude that heat transport is concentrated in the faster-rising parts of plumes. A plausible explanation is that these faster-rising parts of up-plumes lie closer to the roots of the thermals where vertical velocities are greatest. 

The doubly-mixed nature of the scale $\lambda^{1/4}z^{3/4}$ indicates that the flux events themselves depends on both the $\lambda^{1/2}z^{1/2}$-scale eddies organizing and driving the plumes and the $z^{1/2}$-scale length scale of the up-plumes themselves. A $z^{1/2}$ dependence for the up-plumes is consistent with the $\lambda^{1/2}z^{1/2}$ length scale of the $T$ spectra, and indeed with the $z^{-1/2}$ dependence of the cross-sectional areas of the p-plumes as indicated by the mean temperature profile. Overall, the length scale depends on $z^{-3/4}$ because $z$ plays two roles: one as it relates to the cross-sectional areas of the plumes, and the other as it related to what size of impinging outer Richardson eddy is associated with the greatest vertical velocities at each height. We know that there is an optimum size for these impinging eddies because their kinetic energies increase with eddy size, according to the Kolmogorov law, but increased eddy size also means that blocking by the ground reduces the fraction of this energy allocated to the vertical motions at small heights.

\subsection{Small and large wavenumbers}

The mid-range plumes aggregate as height increases, eventually to combine into the $\lambda$-scale clusters we know as thermals. Between these lie cooler areas of descending air with smaller temperature fluctuations, and these areas increase with height. As a result the Fourier representations of  temperature signals change with height, with more variance in the small-wavenumber range but less in the mid-wavenumber range as height increases \citep{McN07,Cho19}. The same is true of the $wT$ covariance in  cospectra, so heat flux is transferred from the mid wavenumber to the small-wavenumber ranges as height increases. The small-wavenumber parts of spectra and cospectra collapse when wavenumbers are scaled on $\lambda$ (Table \ref{ASFL}). 

The mean temperature gradient is insensitive to these changes in spectral representation. The same original plumes still taper in the same way, so the combined area of these up-plumes continue to follow the $z^{-1/2}$ power law, even while their spectral representation is sensitive to their changing aggregation properties. In the small-wavenumber range the covariance increases in proportion to the $(z/\lambda)^{1/2}$ increase at each wavenumber, which $1/2$ power suggests a link to the $(z/\lambda)^{-1/2}$ decrease in the envelope areas of clusters of original up-plumes. We have no formal analysis to confirm this connection.

The large-wavenumber ranges of spectra and cospectra reflect the action of detached outer Richardson eddies on the mid-range plumes, as indicated by the doubly-mixed length scale of the $T$ spectrum in this range. This combines the $\lambda^{1/2}z^{1/2}$ length scale of the plumes with the $z$ length scale of the outer Richardson eddies. They carry this scale even while detached because their population still reflects the influence of the ground. The Richardson cascade is itself scale-free, and characterized by the $-5/3$ power-law Komogorov spectrum. However, this spectrum is also characterized by the absence of eddies with heights significantly greater than $z$, because such large eddies increasingly interact with the ground (become attached) and so behave differently. Detached outer Richardson eddies create fine structure in plumes by stretching and folding them repeatedly, especially at their margins, eventually creating plume filaments at scales where molecular dissipation becomes important. 

Detached outer Richardson eddies have no intrinsic sense of up and down, so the direction of heat transport is set by the mean temperature gradient. Such transport does not change the temperature profile itself. We have a scale-free population of detached Richardson eddies acting on a scale-free temperature profile, so the resulting heat flux is scale free, and so independent of height. This independence is confirmed by observations, which show that $wT$ cospectra collapse over a range of heights when scaled on $H$ (Table \ref{ASFL}). With no flux divergence the mean temperature profile is unaffected. This model of random transport down a pre-existing temperature gradient is consistent with the `eddy diffusion' model of \citet{Wyn72}, which predicts that the $wT$ cospectrum should follow a $-4/3$ power-law in this range, as is observed \citep[e.g.][]{Cho19}.

\section{Plumes within the SFL}

Plumes within the SFL are more complicated. We know that the power-law form of the mean temperature gradient extends down into the SFL, but follows a log law very near the ground \citep{Kad90}. We therefore do not expect plumes to have a single, self-similar form throughout the SFL, but one that changes with height. The nature of that change is our focus here.

Though many experiments have been conducted within the SFL, most published results are of limited value for our work. Only the results of \citet{Lau09}, have been reported using non-dimensionalizations based on our preferred parameter set, \{$\lambda , z,\epsilon_o, u_\epsilon, H$\}; and even those results  are problematic because they were from experiments not originally designed for our kind of analysis. Here we base our interpretations on results from a new analysis of the SLTEST data. We first describe how data were selected, then introduce the new results as we go along.

\subsection{Data selection and processing}

A difficulty encountered when using our parameter set is that calculation of the outer dissipation rate, $\epsilon_o$ must be based on observations made well above the SFL, often at many tens of meters above ground, while the dissipation velocity, $u_\epsilon$ must be based on observations made very near the ground, at small $z/z_s$. That is, instruments must be mounted at two widely-spaced heights if both parameters are to be determined directly. The SLTEST experiment was designed with this in mind, and instruments were arranged so that they would span the SFL on many occasions. \citet{Lau09} used a limited range of instrument heights---just one at their principal site---so $z/z_s$ was essentially an inverse scale variable in their analysis, while $z_s$ was calculated using an empirical relationship for $\epsilon_o$ taken from KY90. We have repeated their analysis using SLTEST data. 

For the present analysis we first selected all daytime runs from SLTEST where wind was from Northerly directions, and for each of these we calculated trial values of $\epsilon_o$ and $u_\epsilon$ by assuming that the highest sonic anemometer lay well above the top of the SFL and the lowest very near its bottom, and so calculated trial values of $z_s$ for each run. Runs were accepted for further analysis when 7.1~m $<z_s<12.8$ m, which values of $z_s$ validate our assumptions that the top sonic anemometer was at $z/z_s>2$ and the bottom one at $z/z_s<0.2$. By this means 32 runs were selected for further analysis, and for each of these the sonic anemometers were well spread across the SFL. This makes $z/z_s$ a true scaled height variable in our analysis. A limitation of the SLTEST results is that even an 18-fold range of anemometer heights is rather small, so we have a rather narrow range of $z_s$ values to work with. While some of the SLTEST results differ from those reported by \citet{Lau09} some also agree, as we point out. The earlier results are still very valuable, being from two experiments where $z_s$ varied widely even while having a limited range of $z_s$ values.

We present our spectral results on wavenumber axes rather than frequency axes even while the basic data are from time-series observations made by fixed instruments. This choice is to facilitate interpretation in terms of the geometries of eddies and plumes. We adopt Taylor's frozen  turbulence hypothesis and use mean wind speed at each level to make the conversion from frequency to wavenumber, in the standard way. In earlier papers we performed similar analyses of specta above the SFL, and for these we used the single mean velocity at the top of the mast to make the conversion at all levels, noting that wind speed varies only slowly with height above the SFL \citep{McN07,Cho19}. Neither procedure is ideal since plumes of different sizes move at different speeds in sheared atmospheric surface layers \citep{Dav74,Che17}. Use of Taylor's hypothesis leads to compression of the wavenumber axis of the $T$ spectra and $wT$ cospectra. Even so, the scales that collapse these spectra and cospectra are unaffected, and so too are our interpretations of those scales.

\begin{figure}[tb]
\vspace{2mm}
	\begin{center}
		\includegraphics[width=.47\textwidth]{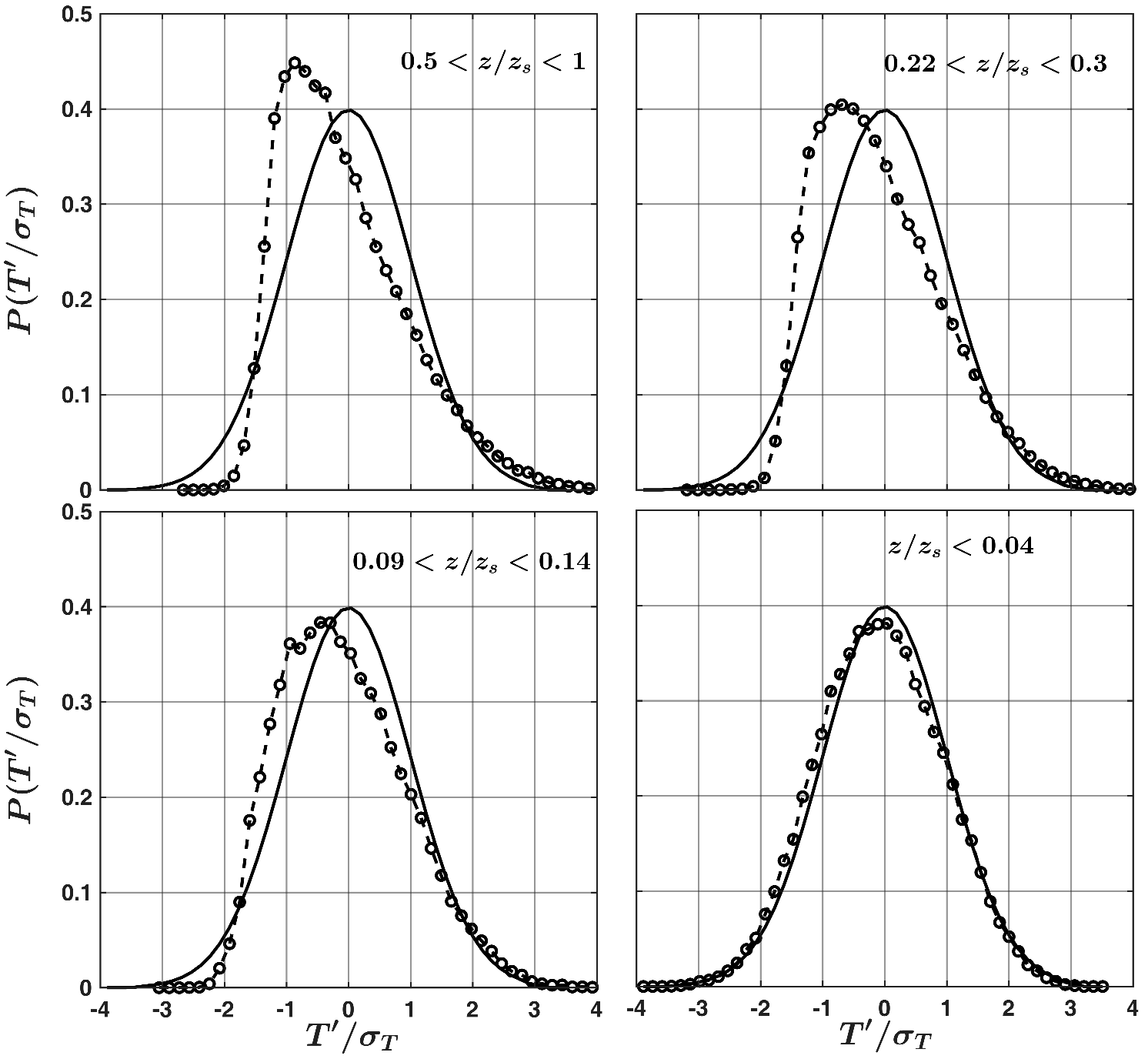}
		\caption{Temperature probability distributions at four heights at SLTEST. Temperature signals were first filtered using a Fourier low-pass filter set at $\kappa z=1$ to remove high-frequency fluctuations. Data are from the 32 runs whose selection is described in the text, except that $\epsilon_o$, and so $z_s$ could not be measured directly for runs where $z/z_s<0.04$. Runs for this case were selected by assuming $z_s=-L$, and further selected so that $-L<100$ m to ensure that the flow regime was always CBL.}
		\label{T_quad}
	\end{center}
\end{figure}

\subsection{$T$ and $wT$ probability distributions}

\begin{figure}[tb]
\vspace{2mm}
	\begin{center}
		\includegraphics[width=.47\textwidth]{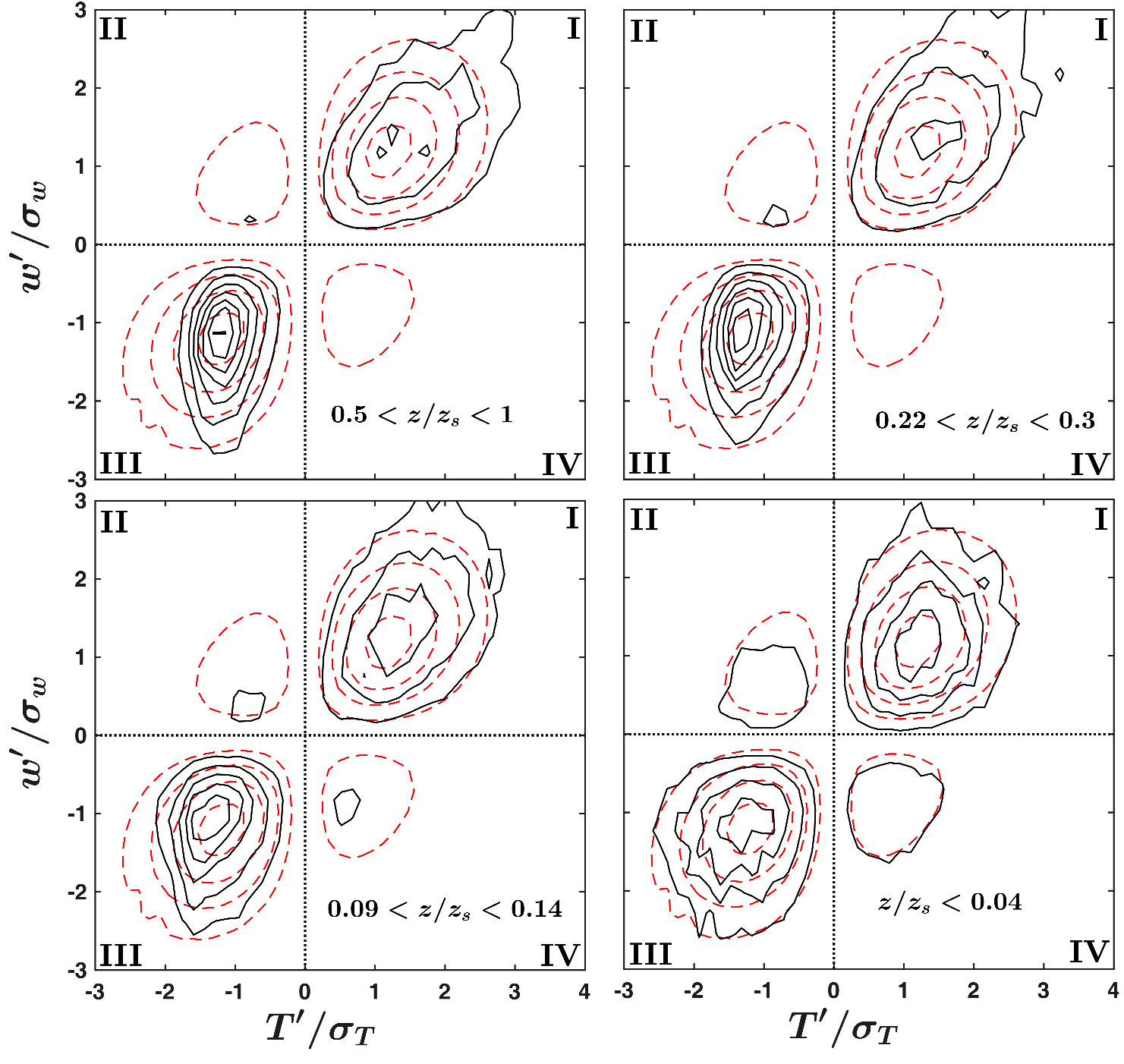}
		\caption{Accumulated heat fluxes, $w'T'/H$ mapped onto the ($w'/\sigma_w$, $T'/\sigma_T$) plane. The red contours represent two Gaussian signals with the same flux. Run details are as for Fig.~\ref{T_quad}.}
		\label{wT_quad}
	\end{center}
\end{figure}

Fig.~\ref{T_quad} shows the $T'$ probability distribution at SLTEST at four height ranges, calculated for just the fluctuations in the mid-range of the spectrum, obtained by first low-pass-filtering the signal with cut-off wavenumber $\kappa z =1$.  Fig.~\ref{wT_quad} shows the joint distribution of the mid-range heat flux mapped on $w'/\sigma_w$ and $T'/\sigma_T$ axes. Starting with Figs.~\ref{T_pdf_largez} and \ref{wT_flux_largez}, which show results from above the SFL, then moving progressively down through the SFL in Figs.~\ref{T_quad} and \ref{wT_quad}, we see a continuous progression from well-organized transport above the SFL to near-Gaussian behavior at its bottom. It seems that plume temperatures have a Gaussian distribution at their source at the ground, in strong contrast to the 2T model, and that this initial distribution is generally maintained as the up-plumes rise. The temperature distribution of the down-plumes can be interpreted as reflecting the combination of a decreasing area fraction of original down-plume air, and its mixing with up-plumes having a Gaussian temperature distribution, with the vertical velocities of the composites becoming increasingly random as height diminishes. (We recall that the up- and down-plume terminology is based on the origin of the plume, not the current direction of movement.) This progression of plume properties is generally consistent with expectation based on the mean profiles, but to go further we must assemble more evidence. We now turn to the $T$ spectra and $wT$ cospectra.

\subsection{$T$ spectra and $wT$ cospectra within the SFL}

\begin{figure}[tb]
\vspace{2mm}
	\begin{center}
		\includegraphics[width=.47\textwidth]{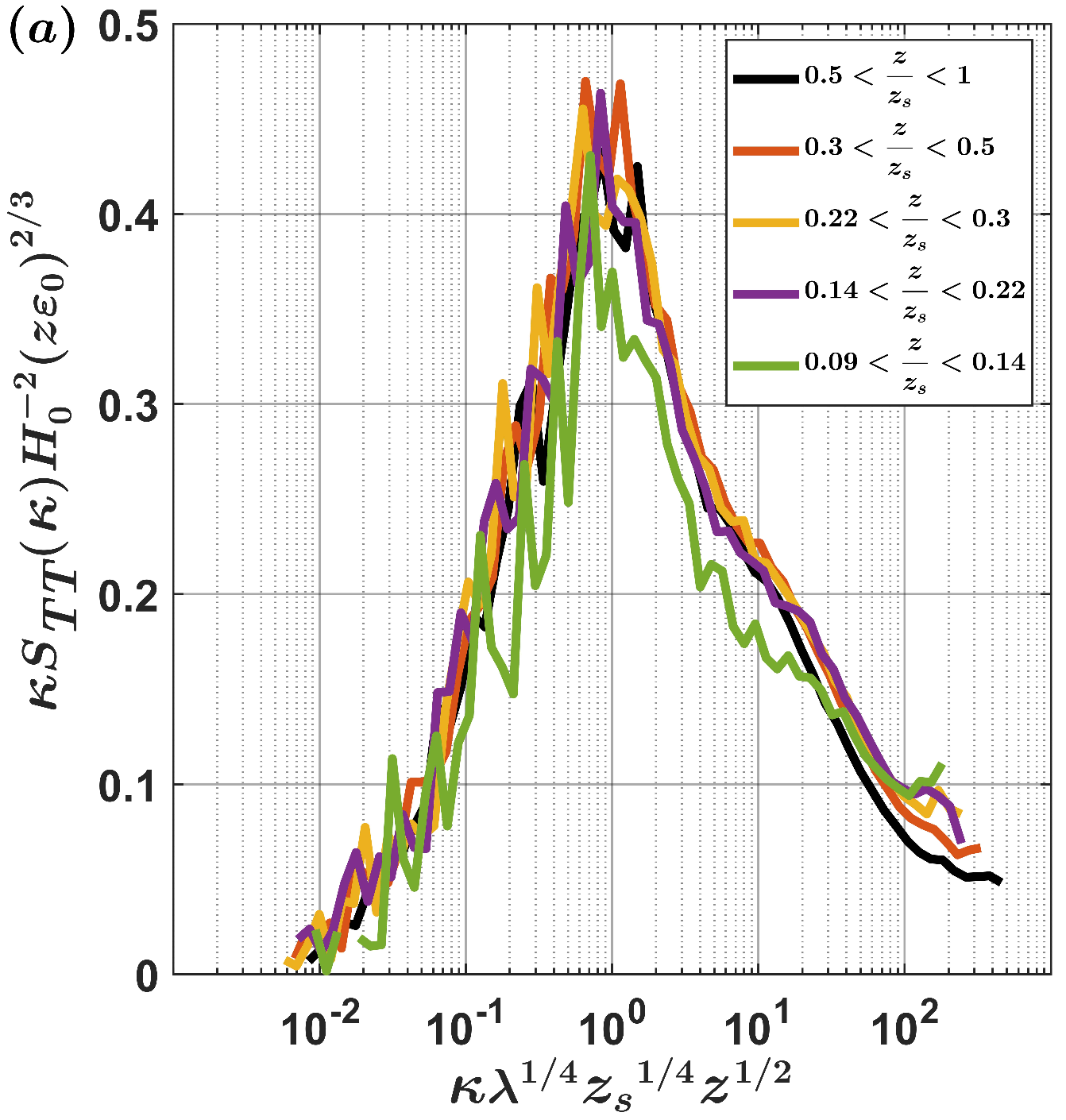}
		\caption{$T$ spectra from within the SFL at the SLTEST site. The variance scale is that which collapses peak heights above the SFL, though the length scale is now doubly-mixed. The scaled spectrum begins to change only at the lowest levels, $z/z_s<0.14$.}
		\label{T_spec_WSFL}
	\end{center}
\end{figure}

\begin{figure*}[tb]
\vspace{2mm}
	\begin{center}
		\includegraphics[width=.9\textwidth]{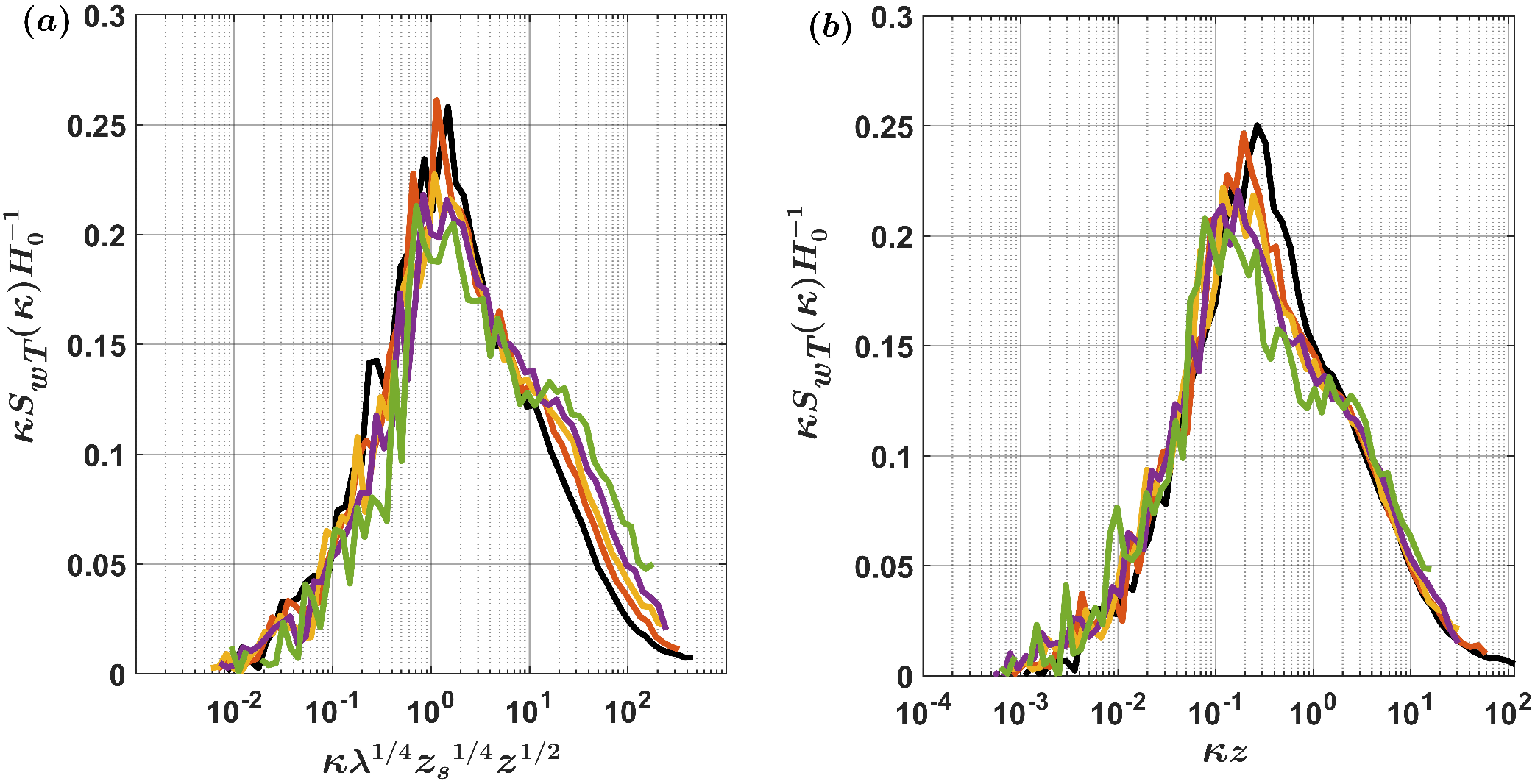}
		\caption{$wT$ cospectra from within the SFL at the SLTEST site. The doubly-mixed plume length scale in (a) is chosen to collapse the positions of the main peaks, while in (b) the simple mixed scale $z$ is chosen to collapse the position of the emerging minor peak at larger wavenumbers.}
		\label{wT_cosp_WSFL}
	\end{center}
\end{figure*}

$T$ spectra for the 32 selected runs from SLTEST are shown in Fig.~\ref{T_spec_WSFL}.   The peak positions collapse when lengths are scaled on the doubly-mixed scale $\lambda^{1/4}z_s^{1/4}z^{1/2}$, in agreement with the collapse shown in \citet[][Fig.~10]{Lau09}. This behavior is therefore consistent across three experimental sites. Fig.~\ref{T_spec_WSFL} also shows that spectrum heights collapse at all wavenumbers in most of the SFL when variance is scaled on $H^2(z\epsilon_o)^{-2/3}$, in agreement with the $z^{1/3}$ power law for the $\sigma_\theta$ profile described above. The same scale collapsed the mid ranges of $T$ spectra above the SFL (Table~\ref{ASFL}). Only the lowest level shown in Fig.~\ref{T_spec_WSFL}, for $0.09<z/z_s<0.14$, deviates significantly. That is, plume self-similarity does not extend right down to the ground.

We note that plume length scales are different above and within the SFL: the appropriate scale above the SFL being the mixed scale $\lambda^{1/2}z^{1/2}$ (Table~\ref{ASFL}), but the doubly-mixed length scale $\lambda^{1/4}z_s^{1/4}z^{1/2}$ within it. These scales both depend on $z^{1/2}$ so the profiles can match at and match at $z=z_s$ without any break in slope. We have already interpreted the mixed scale above the SFL as reflecting the scale of plumes created by the action of $z$-scale impinging outer Richardson eddies. The consistent explanation for the doubly-mixed scale is that we have plumes created by the action of $z$-scale impinging outer Richardson eddies, but embedded within and shaped by the converging flow field of the larger and more powerful, $\lambda^{1/2} z_s^{1/2}$-scale eddies. If so the impinging outer Richardson eddies that act on the $z$-scale plumes have a fixed length at all heights within the SFL. This is consistent with our conceptual model of the SFL, in which the only impinging outer Richardson eddies able to penetrate the SFL are those large enough to survive direct interaction with the more-powerful shear eddies within the SFL.
 
An interesting feature of our results is the small shoulders lying a decade and more to the right of the main peaks in both $T$ spectra and, more prominently, in the $wT$ cospectra shown in Fig. \ref{wT_cosp_WSFL}. These shoulders indicate emerging peaks that become more prominent near the ground and lie at the beginnings of the large-wavenumber parts of the $wT$ cospectra. Fig.~\ref{wT_cosp_WSFL} shows the same cospectra in both panels, but with wavenumbers scaled in two different ways. The doubly-mixed length scale $\lambda^{1/4}z_s^{1/4}z^{1/2}$ gives better collapse of the positions of the main peaks, but simple $z$ scaling gives better collapse of the subsidiary peaks.  At larger wavenumbers the whole $wT$ cospectrum scales on $z$. We also notice that the main peak diminishes with height as the minor peak grows. All of these features support our interpretation that two kinds of eddies, and so two kinds of plumes, transport heat within the SFL. These finding agree with those of \citet[][Fig. 12]{Lau09}, so the results are again consistent across sites.

We note that \citet{Sme07} also reported $wT$ cospectra with two peaks, both in positions consistent with our peaks, but at small $-z/L$ their larger-wavenumber peak is much bigger than that shown in Fig.~\ref{wT_cosp_WSFL}. They interpret its emergence as indicating a transition to a distinct Unstable-Very-Close-to-Neutral (UVCN) flow regime. We agree, but note that our runs are selected so as to avoid sampling this near-neutral regime. The defining characteristic of a CBL regime is the presence of an outer layer deep enough to support an outer Richardson cascade. The thickness of this outer layer reduces as the heat flux decreases because the SFL deepens as neutrality is approached. The remaining outer layer, of depth $(z_i-z_s)$, must eventually become too shallow to accommodate an outer Richardson cascade, whereupon the CBL regime will transition to a near-neutral-boundary-layer (NNBL) regime. We use NNBL rather than UVCN to describe this regime since there is evidence that it persists into slightly stable conditions, where it is described as the `strong turbulence regime' of stable boundary layers. \citet{Sme07} had no direct way to discriminate CBL and NNBL regimes, so their results reflect a mixture of both.  \citet{Lau09} found a similar $z$-scaled peak emerged at small $-z/L$, but their observation height was fixed so this always occurred in near-neutral conditions, and NNBL results were not excluded.

\subsection{Plumes at the bottom of the SFL}

The parameterization of CBLs adopted in this paper is based on an idea introduced by \citet{McN04}. It is that there exists a special kind of eddies, called shear eddies, whose fundamental role is to transport momentum to the ground at small heights where blocking by the ground limits the vertical motions of, and so the transport effectiveness of, larger eddies. Near the ground these shear eddies comprise a self-similar population of attached eddies whose heights scale on $z$. Shear eddies transport momentum down to the ground in stages by the action of ever-smaller shear eddies. That this process is rather inefficient is shown by the high dissipation rates observed near the ground.    

The surface shear stress is steady in laboratory boundary layers, so the dissipation rate can be written as $\epsilon=u_*^3/k$, where $k$ is the von K\'arm\'an constant. In CBLs the surface shear stress is variable under the action of energetic large eddies, which \citet{Tow61} called `inactive'  turbulence, and this variability enhances the dissipation rate within the SFL. We have defined a new velocity scale, $u_\epsilon$, to account for this based on the energy of the shear eddies rather than their ability to transfer momentum in the streamwise direction. Thus $\epsilon=u_\epsilon^3/kz$ for $z<<z_s$. We expect this `dissipation velocity' to be the appropriate scale velocity for scalar transport very near the ground.

Figs~\ref{T_spec_WSFL} and \ref{wT_cosp_WSFL} show spectra in the layer where shear eddies dominate scalar transport, which layer we characterize here as the `bottom of the SFL'. It corresponds to the `dynamic sublayer' in the three-sublayer scheme for the surface layer proposed by KH90. It is a very shallow layer, occupying only the bottom few percent of the SFL. The 32 runs we selected for analysis when preparing Figs~\ref{T_spec_WSFL} and \ref{wT_cosp_WSFL} do not provide adequate information in this layer, so we do not confirm directly that the amplitude of the temperature spectrum scales on $H^2 u_\epsilon^{-2}$. Instead we present qualitative arguments for a model that is consistent with observations from the literature, and with the conceptual model described in the sections above. 

We draw a firm distinction between shear eddies themselves and the plumes they create. Shear eddies are ``patterns of motion'', characterized by their ability to transport momentum. They are highly-organized if transient patterns whose form is dictated by their function. They are also dynamical structures in which motion can be transferred from one part to another, and from one eddy to another, by pressure forces. Plumes, on the other hand, are ``patterns of scalar concentration'' where identities are permanently attached to particular parcels of air. There can therefore be no ongoing association between eddies and plumes, which is to say Reynolds' analogy is a false analogy. Momentum and heat are not transported by the same eddies. Why then do mean velocity and mean temperature profiles both have the same logarithmic form near the ground?

Near the ground plumes see the velocity field of the shear eddies as an un-coordinated churning action, and this action depends on the $z$ and $u_\epsilon$ scales of the shear eddies. The churning will effectively randomize the forms of plumes and their spatial distribution, while imparting to them the $z$ and $u_\epsilon$ scales of the shear eddies. Wind velocity displays a log profile because the shear eddies comprise a self-similar population of attached (scaling on $z$) plumes. The plumes have similar self-similar properties, but with the difference that their self-similarity is inherent in the randomness of their forms, not in their order. We then have
\begin{equation}
\frac{u_\epsilon z}{H}\frac{\partial \overline{\theta}}{\partial z}=\frac{1}{k_\theta}
\label{kthetadef}
\end{equation}
where $k_\theta$ is the temperature counterpart of the von K\'arm\'an constant for the wind profile. In comparison, the standard von K\'arm\'an constant is defined by
\begin{equation}
\frac{z}{u_* }\frac{\partial \overline{u}}{\partial z}=\frac{1}{k}
\label{kdef}
\end{equation}
We write these equations using overbars to indicate time averages because the values of $k$ and $k_\theta$ are almost always calculated using time-series observations made by fixed instruments. It is not possible to compare values of $k_\theta$ and $k$ quantitatively because $u_\epsilon$ has not been used to scale temperature profiles in published studies, but we do not expect $k_\theta$ to be equal to $k$ since the underlying transport mechanisms are different. Different constants have been reported, but all such reports have used scaling procedures different to ours. A recent example is the study by \citet{Che20}.

Both observation and mechanistic arguments therefore agree that plumes have a Gaussian distribution of temperatures at source, even as they emerge from the log layer, and in this respect real up-plumes are quite unlike the single-temperature up-plumes of the 2T model. Even so, and like in the 2T model, their initial temperature signature seems to be maintained as they rise subsequently. This is evident in the slowly-changing distribution of temperatures in quadrant I of Fig.~\ref{wT_quad}. The remarkable thing is that the transition from randomness to organized up-plumes with predominantly upwards local vertical velocity occurs at such a low level, with the transition from randomness starting at $z/z_s\approx 0.04$, or even less, and well progressed by $z/z_s=0.1$. It is this early transition in transport mechanism that answers the question why the log layer for scalars is so much shallower for temperature than for velocity \citep{Bra95}.

\section{General Discussion}

This paper has given an account of the scaling properties of temperature profiles and plumes, and of $T$ spectra and $wT$ cospectra in terms of known properties of the eddies and plumes that comprise turbulent flow in the surface layer of convective boundary layers. The 2T model has contributed to this by allowing us to talk of the cross-sectional areas of plumes, even if only by analogy with idealised plumes defined in a system where such areas can be strictly defined. The distinction between $\sigma_{<w>}$  and $\sigma_w$ is important in this interpretation, particularly within the upper SFL where $\sigma_w$ scales on $u_\epsilon^2$ while $\sigma_{<w>}$ has no simple scale because shear and impinging outer Richardson eddies both contribute to the velocities of composite plumes there.

\subsection{The role of buoyancy}

Our account is based on concepts that are, in some respects, foreign to the statistical fluid mechanics (SFM) understanding of turbulence in CBLs. In particular, buoyancy plays a very different role in our conceptual model of convection. To explain how such a difference can be possible we first review the origins of the SFM model, which model encompasses the work of \citet{Rey95} and \citet{Ric20}, who laid down the concepts, \citet{Mon54} who formalized them in a useful way, and many later authors who have developed empirical relationships based on these concepts.

The origins of SFM lie in the work of \citet{Rey95} who, given the intractability of the Navier-Stokes equations, decided that turbulence could only be addressed statistically. He developed the set of Reynolds-averaged Navier-Stokes (RANS) equations, but formal rigour in this development was achieved much later, by Kolmogorov. The problem was that Reynolds' volume averages, also taken to be averages over large populations of eddies, could not properly be formed in flows whose largest eddies span the flow \citep{McN12}. One can not then define elementary things like mean velocity gradients in terms of Reynolds' volume-averaged variables. Kolmogorov solved this by redefining the volume averages as ensemble averages, but in this formulation the RANS equations can contain no information on the dynamics of flows. To form an ensemble average one takes a series of `snapshots' of a flow, either from separate flow realizations (experiments or simulations) or from a single flow but separated widely enough in time to ensure no statistical connection between the snapshots. This very procedure denies all dynamical interpretations since dynamics is about causal connections between successive states of the flow. Ensemble averaging also smears out any transient spatial patterns in the flow so that the forms of eddies and plumes can not be represented by ensemble-averages. Much recent work has attempted to re-introduce information on flow structures by melding ideas from SFM with ideas taken from the study of physical flow structures, but we have taken an independent path based only on ideas taken from the study of complex dynamical systems (CDS).

Our conceptual departure from SFM is most noticeable in our treatment of buoyancy. \citet{Ric20} extended the RANS equations by adding a buoyancy term to Reynolds' original equations for neutral flows. But Richardson explicitly excluded large eddies from his system and, with the eddy diffusion model in mind, adopted the only alternative and assumed that the energy was transferred to the small, attached eddies associated with the temperature gradient near the ground. His `buoyant production' was therefore a local phenomenon, with its action located near the ground where temperature gradients are largest. This interpretation has persisted \citep[e.g.][]{Mon71,Wyn10} despite the impossibility of any such dynamical interpretation of individual, ensemble-averaged terms in the RANS equations. The RANS energy equation, of which Richardson is part author, simply expresses a balance between the divergence of the flux of mechanical energy and the local dissipation rate, and it can be derived from that principle alone, without need to interpret the dynamical origins of particular terms. Though the magnitude of the `buoyant production' term in the RANS energy equation can be significant, there is no argument to locate this `production' in the smaller eddies. 

\begin{figure}[tb]
\vspace{2mm}
	\begin{center}
		\includegraphics[width=.47\textwidth]{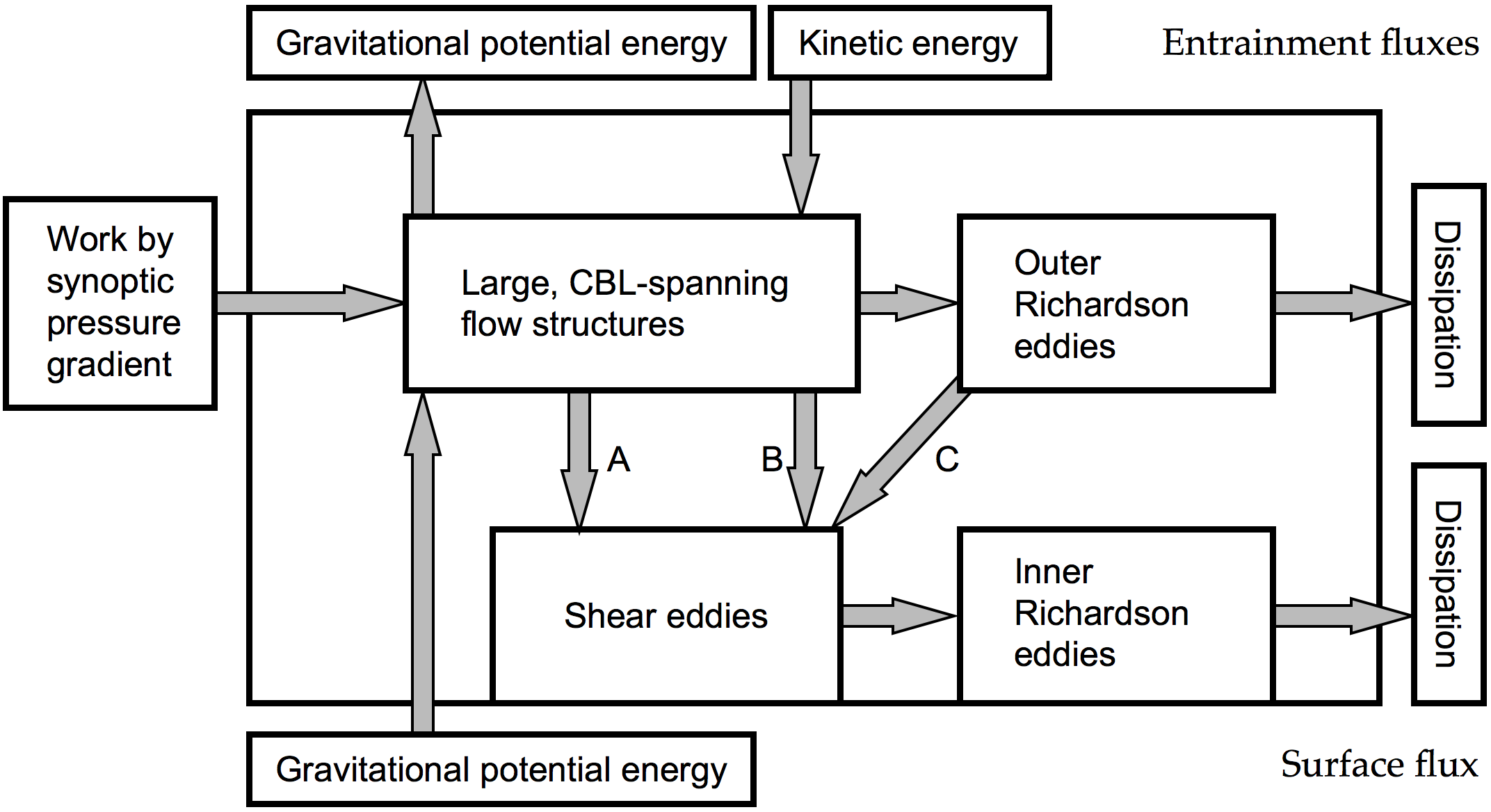}
		\caption{The flow of mechanical energy in CBLs. Arrows represent energy transfers from one kind of eddy to the next. The energy flowing from the large, CBL spanning structures to the shear eddies has been divided into three streams, A, B and C. The first of these supports the mean transfer of shear stress to the ground while the second and third supports the fluctuations. The relative sizes of these flows change with instability. Adapted from \citet{McN12}.}
		\label{MEflow}
	\end{center}
\end{figure}

In CBLs mechanical energy flows into the system at the largest scale and passes down progressively, from one kind of flow structure to the next, to finally be dissipated as heat at the smallest scale, as shown in  Fig.~\ref{MEflow}. The largest flow structures have the greatest energy densities and the longest lifetimes, so smaller eddies and smaller plumes are simply swept along within them, moving in whatever direction and at whatever velocity the larger eddies dictate. In particular, plumes do not accelerate upwards under the action of local buoyancy forces. It is, however, intrinsic to the system-wide organization of the flow that the largest eddies accommodate the buoyancy of smaller plumes by aggregating them and incorporating them into the largest-scale patterns in the flow. We can then best understand the role of buoyancy in the context of all of the other inputs of mechanical energy into the CBL system.

\subsection{Flow of mechanical energy in CBLs}

Fig.~\ref{MEflow} is a schematic diagram of the flows of mechanical energy in the CBL system. It is based on the diagram drawn by \citet{McN12}, but here the large, semi-permanent coherent structures and the largest turbulent eddies are combined and described as large, CBL-spanning flow structures. This change acknowledges the importance of the changing form of the large eddies with changing instability, as described by \citet{Jay18}. An extra energy flow pathway has been added so the mean and fluctuating effects of these large flow structures can be represented separately. Energy now flows to the shear eddies by three pathways, marked A, B and C in  Fig.~\ref{MEflow}. The first component, A, supports the transfer of mean shear stress to the ground while component B supports the fluctuations in shear stress created by the large coherent structures, and component C supports the smaller-scale fluctuations in surface shear stress created by outer Richardson eddies impinging onto the ground. All three will contribute to dissipation near the ground, and so to the value of $u_\epsilon$. Taken alone, an increase in $(F_A+F_B+F_C)$ will therefore cause an increase in $u_\epsilon$ and so an increase in $z_s$, where the $F$s are the fluxes of mechanical energy. However, we know that the form of the large CBL-spanning flow structures do change with stability---from long roll vortices parallel to the wind in near-neutral conditions to cellular convective cells when the CBL is very unstable---so pathways B and C increase in importance as instability increases. The sum of these two energy flows is just $z_i \epsilon_o$, where $z_i$ is the height of the CBL, since the velocity scale of the large eddies can be parameterized as $(z_i \epsilon_o)^{2/3}$ and $\epsilon_o$ also parameterizes energy flow down the outer Richardson cascade \citep{McN07}. The effect is that the ratio $(F_B+F_C)/(F_A+F_B+F_C)$ increases as the large CBL-spanning flow structures become more cellular, so $z_s$ decreases as this ratio increases. Indeed, the ratio acts as a stability parameter for CBLs. 

In Fig.~\ref{MEflow} the large structures pass energy continuously down to smaller and smaller structures in which the form of the source eddies quickly becomes irrelevant. This leads to universal $T$ spectra for the mid- and large-wavenumber ranges of the smaller plumes, and to the parameterization of their velocity scales in terms of the energy flowing through the eddies that shape them. For this reason the outer dissipation rate, $\epsilon_o$, and the dissipation velocity, $u_\epsilon$, are key parameters in our scaling scheme. This perspective is consistent with the importance of the dissipative flux in the theory of far-from-equilibrium systems \citep{Eva07}. 

\subsection{Stability parameters}

In our work $z/z_s$ plays a role somewhat analogous to the role of $-z/L$ in MOST, and indeed the two are correlated \citep{Cho19}. We can use Fig.~\ref{MEflow} to explore the source of this correlation since both $z_s$ and $L$ can be derived from it. As we have seen $z_s$ is given by (\ref{z_sdef}), where $u_\epsilon$ depends on the mechanical energy dissipated within the SFL, and so on the sum of the energies flowing down pathways A, B, and C. If we were to delete the energy flow pathways B and C then the same equation (\ref{MEflow}) would lead to the Obukhov length, $-L$. Without pathways B and C the dissipation velocity, $u_\epsilon$, becomes $u_*$, and $gH/T_0$ replaces $\epsilon_o$ since all of the gravitational energy must now pass directly to outer dissipation. This is so because buoyancy forces act vertically, so none of their energy can go to supporting the mean momentum flux. With these changes $z_s=-L$. Though this leads us to expect correlation between $z_s$ and $L$, it also leads us to expect that the Monin-Obukhov model will break down in windless convection, when $u_*$ goes to zero. Several adjustments to the standard Monin-Obukhov model have been proposed to avoid this. \citet{Bus73} proposed that $u_*$ should have a lower limit $w_*$, where $w_*$ is the Deardorff convective velocity scale. \citet{Bel94} redefined $L$ using $u_*^2+\beta w_*^2$ in place of $u_*^2$, so adding a proxy for the missing energy flow in all conditions. Others, \citep[e.g.][]{Stu88}, have redefined $u_*^2$ as $(\overline{u'w'}^2+\overline{v'w'}^2)^{1/2}$, which is clearly an \emph{ad hoc} adjustment since an ensemble-averaged $\overline{v'w'}$ must be zero, but in practice this also seems to compensate for the missing energy flow \citep[e.g.][]{Wil08}. The other change---setting $\epsilon_o$ equal to $gH/T_0$---has some empirical support in the work of \citet{Kad90}, but their empirical relationship, equivalent to $\epsilon_o=1.1\;gH/T_0$, implies that the contribution of entrained kinetic energy to $\epsilon_o$ outweighs work done against buoyancy in the capping inversion in moderately and strongly convective conditions. This is disturbing in itself, and it must fail in less convective conditions when gravitational potential energy becomes the minor contributor to the outer budget of mechanical energy. Of the two relationships necessary to support a correlation between $z_s$ and $-L$, we expect the $u_\epsilon=u_*$ approximation to become more accurate in nearer-neutral conditions but the $\epsilon_o=gH/T_0$ approximation to become worse.

\section{Conclusions}

In this paper we have surveyed ensemble-averaged observations of temperature profiles, $T$ spectra and $wT$ cospectra and probability plots from within CBLs, and given interpretations of them based on the conceptual model of eddy and plume processes previously introduced by \citet{McN07,Lau09}. This survey has been wider than the previous accounts, and it includes number of new results and interpretations. In particular, we now include the cross-sectional areas of plumes in our interpretations of empirical results. We have been able to identify the $z^{-1/2}$ power law for the mean temperature gradients above the SFL with the narrowing of rising warm plumes as they rise, and to connect this power law with the half power in the $\lambda^{1/2} z^{1/2}$ mixed length scale that collapses the positions of the peaks of $T$ spectra above the SFL. We have also presented results from a new analysis of $T$ spectra from within the SFL at the SLTEST experimental site. These confirm some of the results reported by \citet{Lau09}, and they support a model where plumes near the ground are effectively randomized by the action of shear eddies, but above that, by $z/z_s \approx 0.1$, these are swept together and given form by the action of $z_s$-scale impinging outer Richardson eddies, in which they become embedded. This process also dictates their areas and so the form of the temperature profile further from the ground. Quadrant plots of $T$ probability distributions and joint $w$-$T$ probability distributions support this interpretation.

A key feature of our survey is that it uses concepts that are consistent with our general understanding of the CBL as a dissipative complex dynamical system. In particular, we identify eddies and plumes as emergent structures that embody the high levels of organization. Emergence is a defining characteristic of all far-from-equilibrium, dissipative systems, whose high level of organization is sustained by a constant flow of energy. This energy takes the form of mechanical energy in dynamical systems, and our similarity scheme is based on the parameters that characterize this energy flow. The same perspective leads to an understanding of buoyancy in CBL flows that is very different to that introduced a century ago,  by \citet{Ric20}.

Even so, the new model raises many questions. We still cannot say why mixed scales always involve $1/2$ powers.  We need a new experiment to characterize the temperature profile using the new similarity parameters. We do not know how to model entrainment in a way that will connect energy flows within the CBL to the synoptic-scale flow. This task of parameterizing entrainment is made more complicated by the expectation that the entrainment rate will depend on both the form and the energy of the large structures within the CBL. These are matters for the future. Even so, we are optimistic that the present approach can be extended usefully to the near-neutral and stable regimes in atmospheric boundary layers.


\end{document}